\documentclass[amsmath,amssymb,twocolumn,longbibliography,superscriptaddress,floatfix]{revtex4-2} 

\usepackage[english]{babel}
\usepackage[utf8x]{inputenc}
\usepackage{hyperref}
\usepackage{graphicx}
\usepackage{textcomp}
\usepackage{nameref}
\usepackage{xargs}   
\usepackage[pdftex,dvipsnames]{xcolor}
\usepackage{siunitx}
\usepackage{float}

\hypersetup{
    colorlinks,
    linkcolor={blue!50!black},
    citecolor={blue!50!black},
    urlcolor={blue!80!black}
}

\newcommand{\SiV}{SiV$^-$}

\begin{document}

\title{Structural and optical properties of micro-diamonds with \SiV{} color centers}

\author{Fabio Isa}
\affiliation{CSIRO Manufacturing, 36 Bradfield road, Lindfield, New South Wales 2070, Australia}
\affiliation{Department of Physics and Astronomy, Macquarie University, New South Wales 2109, Australia}
\affiliation{ARC Centre of Excellence for Engineered Quantum Systems (EQUS), Australia}
\author{Matthew Joliffe}
\affiliation{Department of Physics and Astronomy, Macquarie University, New South Wales 2109, Australia}
\author{Brendan Wouterlood}
\affiliation{Department of Physics and Astronomy, Macquarie University, New South Wales 2109, Australia}
\author{Naomi He Ho}
\affiliation{School of Materials Science and Engineering, UNSW Sydney, High Street, Kensington, New South Wales 2052, Australia}
\author{Thomas Volz}
\affiliation{Department of Physics and Astronomy, Macquarie University, New South Wales 2109, Australia}
\affiliation{ARC Centre of Excellence for Engineered Quantum Systems (EQUS), Australia}
\author{Avi Bendavid}
\affiliation{CSIRO Manufacturing, 36 Bradfield road, Lindfield, New South Wales 2070, Australia}
\affiliation{School of Materials Science and Engineering, UNSW Sydney, High Street, Kensington, New South Wales 2052, Australia}
\author{Lachlan J. Rogers}
\email{lachlan.j.rogers@quantum.diamonds}
\affiliation{Department of Physics and Astronomy, Macquarie University, New South Wales 2109, Australia}
\affiliation{ARC Centre of Excellence for Engineered Quantum Systems (EQUS), Australia}

\begin{abstract}
Isolated, micro-meter sized diamonds are grown by micro-wave plasma chemical vapour deposition technique on Si(001) substrates.
Each diamond is uniquely identified by markers milled in the Si substrate by Ga$^{+}$ focused ion beam. The morphology and micro-grain structure analysis, indicates that the diamonds are icosahedral or bi-crystals.
Icosahedral diamonds have higher (up to $\sigma_\mathrm{h}$ = \SI{2.3}{\giga\pascal}), and wider distribution ($\varDelta\sigma_\mathrm{h}$ = \SI{4.47}{\giga\pascal}) of hydrostatic stress built up at the micro-crystal grain boundaries, compared to the other crystals.    
The number and spectral shape of \SiV{} color centers incorporated in the micro-diamonds is analysed, and estimated by means of temperature dependent photoluminescence measurements, and Montecarlo simulations.
The Montecarlo simulations indicates that the number of \SiV{} color centers is a few thousand per micro-diamond.
\end{abstract}

\maketitle

\section*{Introduction}

Colour centres in diamond have become well established architectures for quantum technologies\cite{thiering2020chapter}.
%
In particular, the negatively-charged silicon vacancy (\SiV{}) centre has proven to be an excellent single photon source \cite{rogers2014multiple, sipahigil2014indistinguishable} and quantum-optical platform \cite{
sipahigil2016integrated,
wan_large-scale_2020, 
schroder_scalable_2017}. 
For many quantum information applications it is important to achieve isolated single defects in fabricated structures and nano-diamonds \cite{
jantzen2016nanodiamonds,  
rogers2019single,         
chen2020building,         
zhang_strongly_2018}.	  
There have been successful demonstrations of single \SiV{} centres in nano-beam waveguides \cite{sipahigil2016integrated}, and bottom-up engineering of complex quantum systems from nano-diamonds has been proposed \cite{rogers2019single}.
However, other applications demand high densities of \SiV{} centres in confined geometries \cite{de_feudis_large-scale_2020, schutz_ensemble-induced_2020}.
The spectral stability and strong zero-phonon line of \SiV{} centres makes them attractive for exploring cooperative effects such as superradiance \cite{bradac2017room-temperature}.
Emerging techniques for optical trapping have applications in particle sorting and inertial sensing, but demand high concentrations of optical dipoles \cite{juan2016cooperatively}.
%

%
%
Here we report a novel approach to fabricate MDs with large ensembles of \SiV{} color centers, and to calculate their number density.
Firstly, isolated and high crystal quality micro-diamonds (MDs), are grown by micro-wave plasma chemical vapour deposition (MWCVD) technique on silicon.
Each MD is uniquely identified and addressed by micrometer-wide markers milled in the Si substrate by focused ion beam (FIB) technique in the scanning electron microscope chamber.
The micro-structure and morphology of individual MDs is investigated by a combination of scanning electron microscopy (SEM), atomic force microscopy (AFM) and polarised confocal Raman spectroscopy.
Finally, the number density and spectral shape of \SiV{} ensembles in MDs is investigated and calculated by temperature dependent photoluminescence spectroscopy, and Montecarlo simulations.
These simulations qualitatively reproduce the ensemble spectra by realistically simulating individual \SiV{} spectra for a given number of \SiV{} centres ranging in axial strain, transverse strain, and temperature.
It is possible to estimate the number of \SiV{} centres to be a few thousand per MD.

\section*{\label{sec:results}Results}

\subsection{\label{sec:growth_morphology}Growth and morphology of micro-diamonds on Si}

Micrometer-sized diamonds, were grown by MWCVD with CH$_{4}$ and H$_{2}$ as precursor gases, directly on Si(001) substrates without any seeding procedure.
The MDs nucleate on the defects of the SiC/Si(001) thin layer which is formed during the CH$_{4}$ plasma process, with a density below 1 mm$^{-2}$. 
The X-ray photoelectron spectroscopy (XPS) analysis (see supporting material) clearly indicates the formation of a SiC layer both in the C 1s and Si 2p signals. 
The atomic concentration of the SiC layer is 26 \% $\pm$ 1 \% calculated from the integral peak intensities (see section \autoref{sec:experimental_methods}), considering that the$nm$ XPS penetration depth ranges between 5 -- \SI{10}{\nano\metre}, it corresponds to a few SiC monolayers.
The formation of a SiC interlayer enables the nucleation, and crystal growth of diamonds without any seeding procedure \cite{Stoner1992,Jiang1994}.  

The growth process (see section \ref{sec:experimental_methods}) has been optimized to form MDs with well defined crystal facets, having the so-called crystal quality (intensity ratio between the Raman signal of C sp$^{3}$ and sp$^{2}$) higher than 99.5\% measured by Raman spectroscopy \cite{Bak2008,Sails1996,Prawer2004}, and a crystal size of about \SI{2}{\micro\metre}. 

\begin{figure*}
	\centering
	\includegraphics[width=1\linewidth]{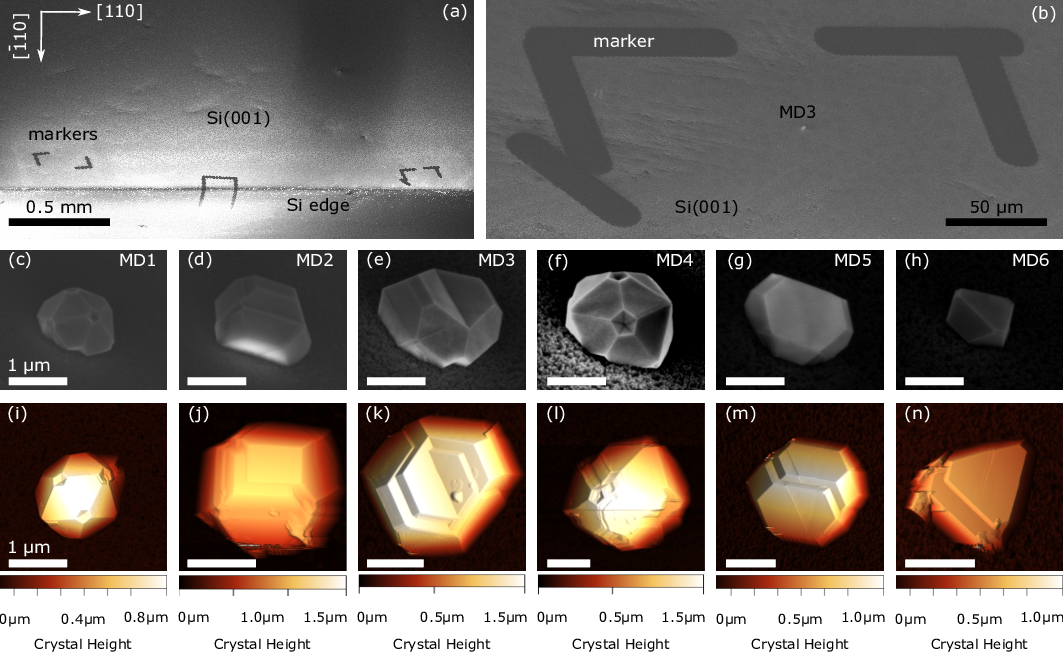}
	\caption{
		(a) SEM image of the Si substrate with 3 different FIB markers.
		(b) Magnification of one FIB marker indicated in (a), MD3 is clearly visible in the centre.
		(c) - (h) Perspective view SEM images of MD1, MD2, MD3, MD4, MD5 and MD6, respectively.
		(i) - (n) AFM images of the same MDs depicted in (c) - (h). 
		In (c) - (n) the white scale bar is \SI{1}{\micro\metre}.
		}
	\label{figure_SEM_AFM}
\end{figure*}

In order to measure and to correlate the structural and optical properties of specific diamond crystals, six different MDs, with various morphology and structure, were identified by SEM and labelled by focused ion beam (FIB) milled markers.
\autoref{figure_SEM_AFM}(a) and (b) report perspective view SEM images of the Si(001) substrate after the MWCVD MD growth, and FIB milling of markers. 
The FIB markers (dark grey) are \SI{2}{\micro\metre} deep and \SI{30}{\micro\metre} wide with unique shapes. These aspects make the markers easily addressable by atomic force microscope (AFM), and confocal optical spectroscopy techniques. 
In \autoref{figure_SEM_AFM}(b), a single MD, labelled as MD3, is clearly visible in the middle of the FIB marked area.

The morphology of six different micro-diamonds, labelled as MD1 - MD6 is investigated by perspective view SEM and AFM, the results are displayed in \autoref{figure_SEM_AFM} (c) - (h), and (i) - (n), respectively. 

%
Both SEM and AFM analysis shows a similar crystal size of $\sim$ \SI{2}{\micro\metre} for every MDs, indicating that all crystals nucleated in a similar stage of the CVD process.
Each MD presents well-defined \{100\} and \{111\} crystal facets, but the overall crystal shape and morphology are different.
Indeed, the diamond crystal shape is determined by the growth velocity parameter $\alpha=\sqrt{3}v_{100}/v_{111}$, where $v_{100}$ and $v_{111}$ are the growth velocity of the \{100\} and \{111\} surfaces, respectively \cite{Buhler2000}. 
If $\alpha \leq 1.5$, the \{111\} facets are the fastest growing, and therefore the crystal will be predominantly \{111\} textured. 
On the contrary, if $\alpha \geq 1.5$ the \{100\} facets are the fastest growing, and therefore the crystal will be predominantly \{100\} textured. 
The parameter $\alpha$ strongly depends on the local plasma and gas mixture conditions, typically it increases with high methane concentrations, and it decreases with high growth temperatures. 
The overall MD shape does not depend on the parameter $\alpha$ only, but also on the formation of multiple crystal twins separated by grain boundaries \cite{Shechtman1993,Shechtman1993a}. 

The SEM and AFM images in  \autoref{figure_SEM_AFM} of MD2, MD3, MD5 and MD6 indicate that the crystals are constituted by two different crystal grains, and they are bounded by \{100\} and \{111\} facets.

Differently, the SEM and AFM images in \autoref{figure_SEM_AFM} (a), (f) and (i), (l) of MD1 and MD4,  indicate that these diamonds are dimpled icosahedral crystals, formed out of \{111\} facets with 5-fold pseudo symmetry axes \cite{Buhler2000,Shechtman2006}.
The five-fold crystal symmetry originates from the formation of five neighbouring $\Sigma3$ twins.
Even tough this crystal structure lacks of 7.33 $^{\circ}$ to a full 360 $^{\circ}$ closure, the mismatch is compensated by internal strain at the grain boundary regions \cite{Shechtman1993}.
If the parameter $\alpha = 1.5$, the five-fold symmetric structure results in a perfect icosahedral crystal. 
As the growth conditions change towards higher values of $\alpha$, dimples form on the top of the symmetry axes.
%

\begin{figure}
	\centering
	\includegraphics[width=1\linewidth]{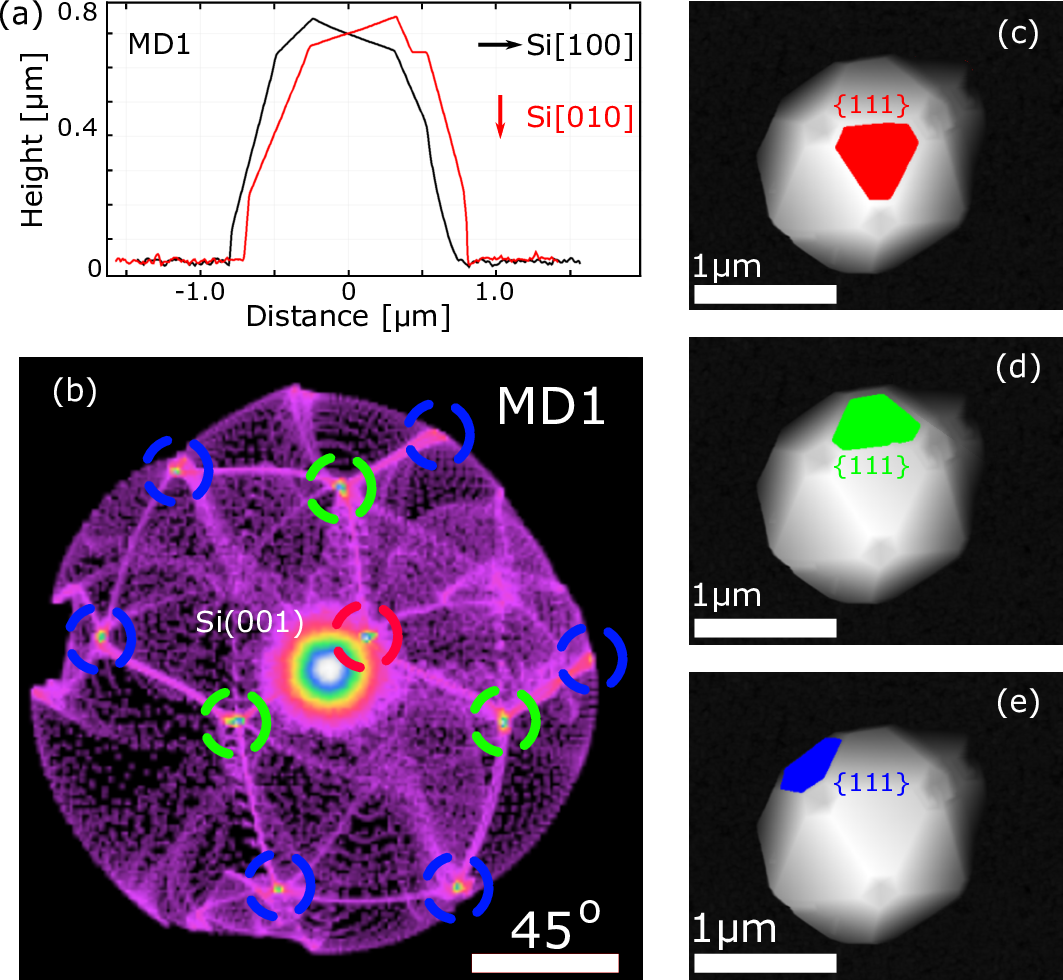}
	\caption{
		(a) Height profile of MD1 taken through the crystal centre along the Si [100] (black) and Si[010] (red) directions.
		(b) Stereographic map of the MD1 crystal facets obtained by AFM analysis.
		The high-intensity central peak is related to the Si(001) substrate.
		The red, green and blue dashed circles indicate the family of \{111\} crystal facets marked in (c) - (e). 
		(c) - (e) AFM images of MD1 with \{111\} facets marked in red, green and blue, according to the stereographic map reported in (b).
		}
	\label{figure_stereographic}
\end{figure}

%
In order to get more insights into the diamonds' micro-structure, the crystal height profile along the two orthogonal Si[100] and Si[010] directions, together with the stereographic map of the crystal facets distribution \cite{Robinson2009}, are extracted from the AFM data for each MD. 
\autoref{figure_stereographic} reports the crystal profile and stereographic map for MD1; all MDs are analysed analogously.
AFM data are flattened with respect to the Si(001) substrate surface.
The crystal height profile of MD1 along the Si[100] (black) and Si[010] (red) directions, is reported in \autoref{figure_stereographic} (a). 
Here, the different \{111\} facets of the icosahedron are clearly visible in both crystal directions.
In the stereographic map of \autoref{figure_stereographic} (b), the bright central spot corresponds to the Si(001) substrate, while the different groups of \{111\} facets forming the icosahedral crystal shape, are marked by: one red dashed circle, three green dashed circles, and six blue dashed circles.  
The corresponding \{111\} crystal facets are marked in red, green and blue in \autoref{figure_stereographic} (c), (d) and (e), respectively.     

\begin{figure}
	\centering
	\includegraphics[width=1\linewidth]{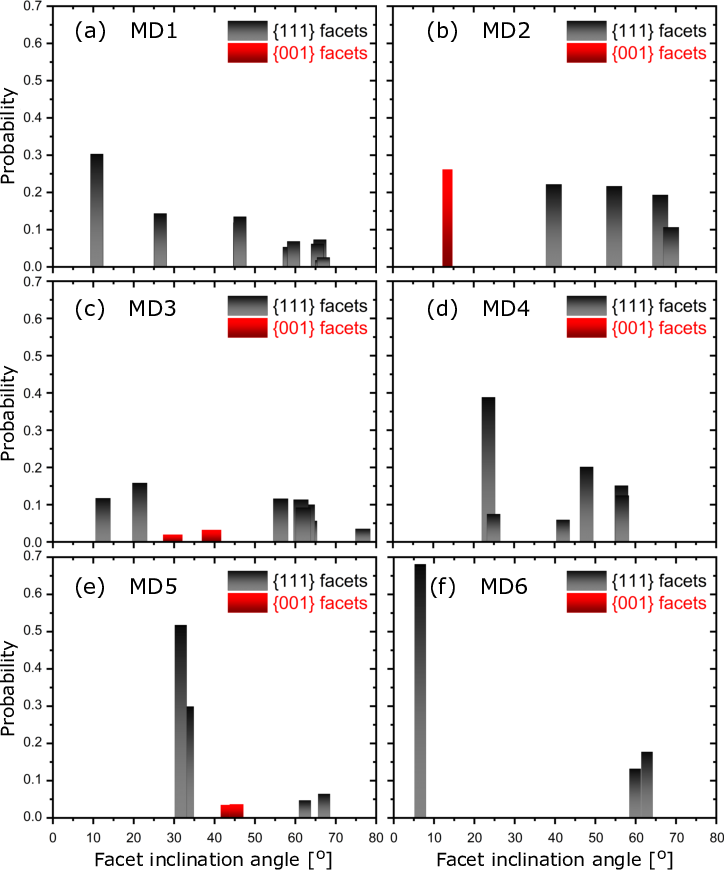}
	\caption{
		(a) - (f) Probability distribution of the (001) (red) and \{111\} (black) facet inclination angle with respect to the Si(001) substrate, for MD1, MD2, MD3, MD4, MD5 and MD6, respectively.
		The Facet inclination angle is measured by calculating the stereographic map from the AFM data as illustrated in figure \ref{figure_stereographic} (b).
		}
	\label{figure_facets_statistics}
\end{figure}

%
A summary of the facet distribution for MD1 - 6, obtained by AFM and stereographic map analysis, is reported in \autoref{figure_facets_statistics} (a) - (f), respectively.
The \{100\} and \{111\} facets are distinguished by the facet symmetry, as explained in reference \cite{Buhler2000}.
In \autoref{figure_facets_statistics}, the facet inclination angle represents how steep the crystal facet is with respect to the Si(001) substrate surface.
This angle is determined by the radial distance between a point and the center in the stereographic map (e.g. see \autoref{figure_stereographic} (b)).
The probability indicates the normalised ratio between the number of points belonging to a \{111\} (black histograms) or \{100\} (red histograms) diamond facet with a specific inclination angle, and the total number of points.
\autoref{figure_facets_statistics} indicates that MD1, MD4 and MD6 present \{111\} facets only, while MD2, MD3 and MD5 exhibit also \{100\} facets.
The \{100\} surface coverage is maximum 30 \% for MD2, thus indicating that the value of the growth parameter is $\alpha \geq$ 1.5.

\subsection{\label{sec:polarised_raman}Polarised Raman Spectroscopy}

%
Confocal Raman spectroscopy is a very powerful technique to analyse the micro-structural properties of diamonds, such as crystal quality, twins formation and lattice stress \cite{Prawer2004}.
Diamond is a face centered cubic Bravais lattice with $O_{h}$ point group symmetry.
The Raman scattering effect in diamond is characterised by a triply degenerate optical phonons belonging to the $F_{2g}$ symmetry group, whose scattering tensors $R_{1,2,3}$ are described by equation \ref{equation_raman_tensors}.  
\begin{equation}\label{equation_raman_tensors}
\begin{split}
R_{1}&=\left[\begin{array}{ccc}
0 & 0 & 0 \\ 
0 & 0 & d \\ 
0 & d & 0
\end{array}\right],
R_{2}=\left[\begin{array}{ccc}
0 & 0 & d \\ 
0 & 0 & 0 \\ 
d & 0 & 0
\end{array}\right],\\
R_{3}&=\left[\begin{array}{ccc}
0 & d & 0 \\ 
d & 0 & 0 \\ 
0 & 0 & 0
\end{array}\right] 
\end{split}
\end{equation}
In equation \ref{equation_raman_tensors}, $d$ represents the Raman tensor element which depends on the crystal electrical polarizability.
The Raman tensors $R_{1}$ and $R_{2}$ are associated to transverse optical (TO) phonon modes along the [100] and [010] crystal directions, respectively.
Differently, $R_{3}$ relates to the longitudinal optical (LO) phonon mode along the [001] crystal direction. 
The overall measured Raman scattering intensity $I_\mathrm{measured}$ is a combination of the three different phonon modes, and it can be expressed by equation \ref{equation_raman_intensity}.
\begin{equation}\label{equation_raman_intensity}
I_\mathrm{measured}\propto\sum_{n=1}^{3}|e_\mathrm{i}R_{n}e_\mathrm{s}|^{2}
\end{equation}
In equation \ref{equation_raman_intensity}, $e_\mathrm{i}$ and $e_\mathrm{s}$ are the linear polarization vectors of the incident laser beam and scattered light, respectively.
Thus, the Raman scattered intensity strongly depends on the relative orientation between the diamond crystal axis and the incident/scattered polarization directions of light.
This enables to distinguish crystal grains with different crystallographic orientations. 

In this work, $e_\mathrm{s}$ is set along the Si substrate [010] crystal direction, denoted as $y$ in the Porto Raman notation.
Differently, the angular rotation $\phi$ of the linearly polarized incident light $e_\mathrm{i}$ is controlled by a rotating $\lambda/2$ waveplate, being $e_\mathrm{i}$=[cos$\phi$ sin$\phi$ 0].
Detailed studies of the angular dependence of the diamond Raman scattering intensity for different crystal facets can be found in references \cite{Prawer2004,Steele2016,Mossbrucker1996}. 
In summary, the Raman intensity for backscattered measurements normal to the diamond (001) and (111) facets is $I_{(001)} = d^{2} cos^{2} (\phi-\phi_\mathrm{max})$, and $I_{(111)} = 2d^{2}/3$, respectively.
The angle $\phi_\mathrm{max}$ indicates the tilt of the investigated crystal with respect to the Si substrate [010] and [100] directions. 
If the incident and scattered beams are not perpendicular to the crystal facets, the Raman scattered intensity dependence on $\phi$ becomes more complicated, and $I_{(111)}$ is function of $\phi$ \cite{Ramabadran2018}.
If the diamond lattice is under hydrostatic stress $\sigma_\mathrm{h}$, the Raman peak position $\nu$ is shifted by $\varDelta\nu$ from the relaxed value of $\nu_{0}$ = 1332 cm$^{-1}$, according to the equation $\nu = \nu_{0} + \varDelta\nu$, where $\sigma_\mathrm{h} = 0.34(\mathrm{GPa/cm}^{-1}) \varDelta\nu$ \cite{Liscia2013, Prawer2004}.

\begin{figure}
	\centering
	\includegraphics[width=1\linewidth]{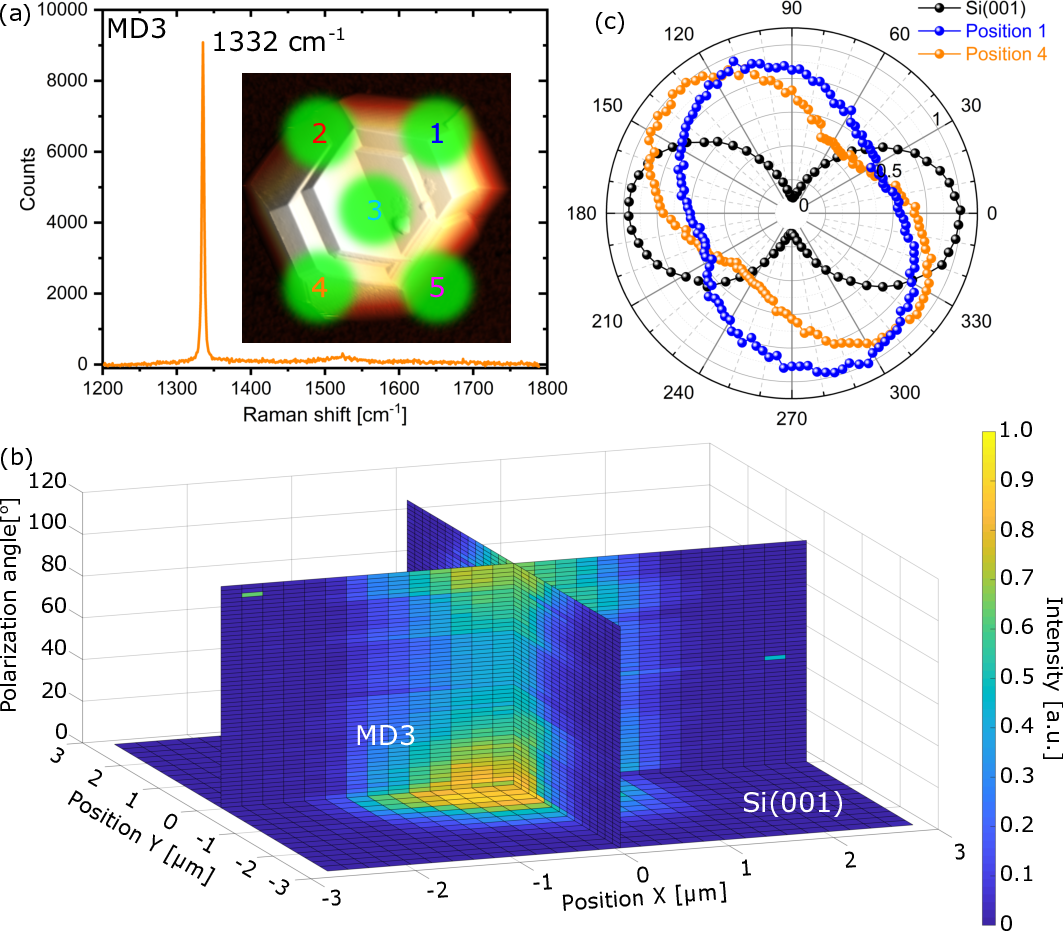}
	\caption{
		(a) Raman spectrum of MD3 measured in position 4 (see inset), the diamond Raman peak around 1332 cm$^{-1}$ is labelled.
		The inset shows the AFM image of MD3 with 5 different laser positions spaced in X-Y directions by \SI{0.6}{\micro\metre}: position 1(blue), position 2(red), position 3 (cyan), position 4 (orange), and position 5 (magenta).
		(b) Colour-coded map of the diamond Raman peak intensity of MD3, sliced trough the crystal centre. 
		The vertical axis represents the angular rotation $\phi$ of the linearly polarized laser light. 
		(c) Polar plot of the Raman peak intensity of the Si(001) substrate (black), diamond at position 1 (blue), and position 4 (orange), as a function of the angular rotation $\phi$ of the laser light linear polarization.
	}
	\label{figure_Raman_Diamond_polarization}
\end{figure}

%
A typical Raman spectrum obtained from the MDs is reported in \autoref{figure_Raman_Diamond_polarization} (a).
It clearly presents the sharp diamond Raman scattering transition around 1332 cm$^{-1}$, while the contribution from amorphous and sp$^{2}$ C at higher wavenumbers is almost negligible, corroborating the good crystal quality obtained by CVD growth.
The micro-structure of the MDs can be revealed by mapping at different X-Y positions the diamond scattered Raman signal as a function of the incident beam light polarization angle $\phi$.  
\autoref{figure_Raman_Diamond_polarization} (b) represents a 3D colour coded map of the diamond Raman peak intensity around 1332 cm$^{-1}$ as a function of the X-Y sample position (in steps of \SI{200}{\nano\metre}), and the angular rotation $\phi$ ($0 \leq \phi \leq 120 ^{\circ}$) of the laser polarization.
The 3D map is sliced along the two orthogonal directions Si[100] and Si[010], through the center of MD3.
The high intensity Raman peak regions comes from scattering of the MD3 crystal, while the dark blue region at low intensities corresponds to the Si(001) substrate (the Raman mode of Si is around 520 cm$^{-1}$ \cite{Parker1967}).
Interestingly, the modulation of the diamond Raman intensity \cite{Prawer2004,Steele2016,Mossbrucker1996} as a function of the laser polarization angle $\phi$ is not uniform throughout the entire MD surface, as it is expected for a single-crystal diamond.

In order to analyse the micro-structure of the MDs, Raman spectra are recorded at different laser light polarizations ($0 \leq \phi \leq 360 ^{\circ}$, in steps of 3$^{\circ}$), in five different diamond positions, labelled in \autoref{figure_Raman_Diamond_polarization} (a) inset as 1 - 5 in blue, red, cyan, orange, and magenta, respectively.
Position 3 is set at the MD center, the other positions are spaced by \SI{600}{\nano\metre} along the X-Y orthogonal directions.
As an example, \autoref{figure_Raman_Diamond_polarization} (c) presents polar plots of the Raman peak intensity for the Si(001) substrate (black) as a reference, and MD3 at two opposite positions: 1 in blue, and 4 in orange.
The Si Raman peak intensity $I_{\mathrm{Si}(001)}$ shows a clear dependence versus the incident light polarization angle $\phi$ according to the equation $I_{\mathrm{Si}(001)} = d^{2} cos^{2} (\phi-\phi_\mathrm{max})$ \cite{Prawer2004}, with its maximum at $\phi_\mathrm{max}$ = 0, along the Si[010] crystal direction.
Differently, the intensity maxima of the diamond Raman peaks for MD3 at position 1 and 4 are offset by $\phi_\mathrm{max}$ = -59$^{\circ}$ $\pm$ 3$^{\circ}$, and -39$^{\circ}$ $\pm$ 3$^{\circ}$ with respect to the Si(001) substrate, respectively.   
This result confirms that MD3 is constituted by diamond grains with different crystallographic orientations.

\begin{figure}
	\centering
	\includegraphics[width=1\linewidth]{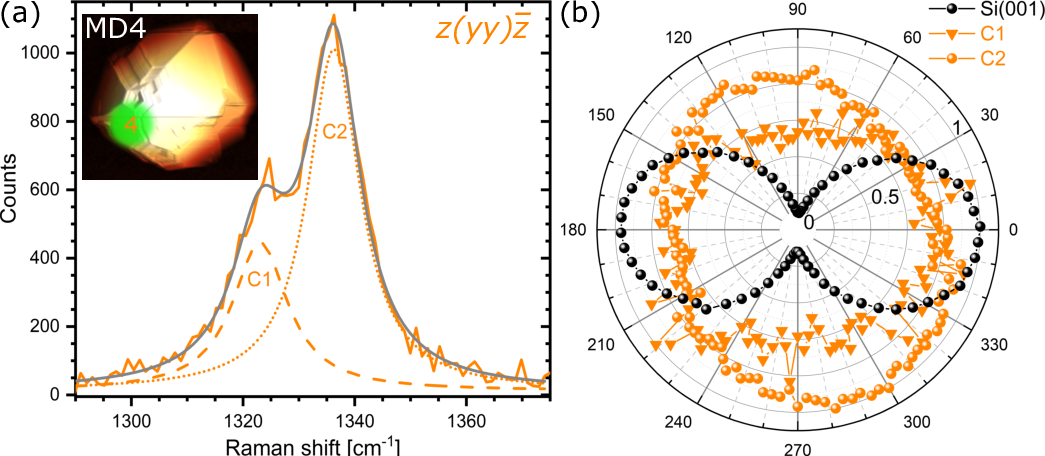}
	\caption{
		(a) Raman spectrum (orange solid line) of MD4 measured at position 4, as indicated in the AFM figure inset. 
		The grey solid line is the bi-Gaussian fit of the Raman signal.
		C1 (dashed orange line), and C2 (dotted orange line) are the Gaussian contributions of two different diamond nano-crystals.
		(b) Polar plot of the Raman peak intensity of the Si(001) substrate (black), MD4 at position 4 Gaussian component C1 (orange triangles), and Gaussian component C2 (orange spheres) as a function of the angular rotation $\phi$ of the laser light linear polarization.
		}
	\label{figure_Raman_Diamond_doublet}
\end{figure}

The micro structural results, obtained from Raman scattering of the icosahedral crystals, are summarized in \autoref{figure_Raman_Diamond_doublet}.
\autoref{figure_Raman_Diamond_doublet} (a) shows the Raman spectrum of MD4 measured at position 4, close to the dimpled corner where 5 different \{111\} facets are joining (see the inset schematic figure).
The diamond Raman signal (orange solid line), recorded in the \textit{z(yy)\={z}} configuration, is clearly split in two distinct components, and it is fitted by the sum of two Gaussian curves (grey solid line).
The first Gaussian component C1 (dashed orange curve) is centered around 1324 cm$^{-1}$, while the second one C2 (dotted orange curve), is centered around 1334 cm$^{-1}$.  
Presumably, the significant difference in Raman peak position between the C1 and C2 crystal components of MD4, is ascribed to large stress built at the five-fold grain boundary \cite{Habka2010} because of lattice mismatch and C sp$^{2}$ defects \cite{VonKaenel1996}.       
Raman peak intensities as a function of the incident laser light polarization angle $\phi$, for the Si(001) substrate, MD4 C1 and C2 crystal components, are reported in \ref{figure_Raman_Diamond_doublet} (b) in black spheres, orange triangles, and orange spheres, respectively.  
The Si Raman peak intensity has a clear polarization dependence as described above, with the maximum along the Si[100] direction.
Differently, the Raman peak intensity of the two diamond crystal components C1 and C2 is almost independent from the incident laser light polarization angle $\phi$.
This result may be attributed to the weak dependence of the Raman signal on the angle $\phi$ for \{111\} crystal facets \cite{Steele2016}, and to the Raman scattering contribution of differently oriented, and distorted crystal grains.

\begin{figure}
	\centering
	\includegraphics[width=1\linewidth]{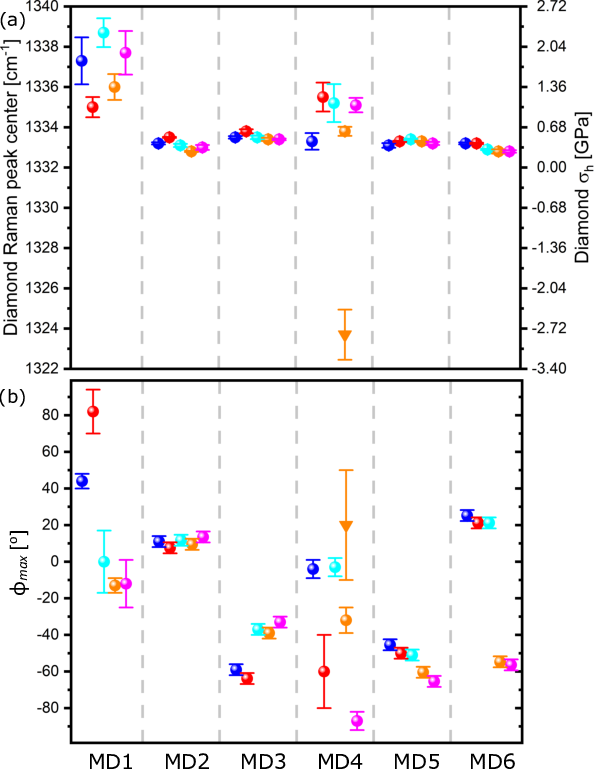}
	\caption{
		Distribution of crystal properties measured by Raman spectroscopy of MD1 - 6 at different positions: 1 (blue), 2 (red), 3 (cyan), 4 (orange), and 5 (magenta). The two Gaussian components of the Raman peak of MD4 at position 4 are labelled by orange triangles (C1), and orange spheres (C2).
		(a) Diamond Raman peak center (left) and calculated Hydrostatic diamond stress (right) $\sigma_\mathrm{h} = 0.34(\mathrm{GPa/cm}^{-1}) \varDelta\nu$ \cite{Liscia2013, Prawer2004}.
		(b) Angular rotation of the laser light polarization $\phi_\mathrm{max}$ giving the maximum Raman scattered intensity.
		}
	\label{figure_Raman_Diamond_statistics}
\end{figure}

Analogously, all MDs have been analysed by Raman spectroscopy with the procedure reported in \autoref{figure_Raman_Diamond_polarization} and \autoref{figure_Raman_Diamond_doublet}.  
A summary of the micro-crystal properties obtained for each MD at position 1 - 5 (marked in different colors as indicated in \autoref{figure_Raman_Diamond_polarization} (a) inset) is summarised in \autoref{figure_Raman_Diamond_statistics}.
Clearly, the crystal properties of the icosahedral diamonds MD1 and MD4 differ from those of the others.
The diamond Raman peak center, see \autoref{figure_Raman_Diamond_statistics} (a), of MD1 and MD4 varies significantly between 1339 to 1324 cm$^{-1}$, while it ranges between 1332.8 and 1333.8 cm$^{-1}$ only, for the other crystals.
Analogously, the calculated hydrostatic diamond stress $\sigma_\mathrm{h}$ follows the same trend, ranging from 2.3 to \SI{-3.3}{\giga\pascal} for MD1 and MD4, and varying between 0.3 and \SI{0.5}{\giga\pascal} only, for the other crystals.
The large difference of diamond Raman peak center and $\sigma_\mathrm{h}$ for icosahedral MD1 and MD4 is ascribed to the significant lattice deformation and high density of defects formed at the multiple grain boundaries regions.

The presence of several crystal grains, with different orientation, causing the large stress in MD1 and MD4 is clearly elucidated in  \autoref{figure_Raman_Diamond_statistics} (b).
Here, the values of $\phi_\mathrm{max}$, measured at different crystal positions, of MD1 and MD4, significantly scatter between 80$^{\circ}$ and -10$^{\circ}$, 20$^{\circ}$ and -87$^{\circ}$, respectively.
The large error bar, and poor spatial consistency of $\phi_\mathrm{max}$ as a function of the different (1 -- 5) crystal positions, indicate that MD1 and MD4 are constituted by several micro-crystals forming highly-stressed and defective grain boundaries. 
Differently, for MD3, MD5 and MD6, the values of $\phi_\mathrm{max}$ at different crystal positions, can be grouped in two distinct sub-ranges, around -60$^{\circ}$ and -40$^{\circ}$, around -60$^{\circ}$ and -50$^{\circ}$, around 20$^{\circ}$ and -55$^{\circ}$, respectively.
The presence of two distinct sub-ranges of $\phi_\mathrm{max}$ values at different crystal positions, indicate that MD3, MD5 and MD6 are formed by two crystal grains with different crystallographic orientations. 
Interestingly, the value of $\phi_\mathrm{max}$ for MD2 is 11$^{\circ}$ $\pm$ 3$^{\circ}$ for every crystal position, suggesting that it is a single diamond crystal.
On the other hand, by  comparing the MD2 $\phi_\mathrm{max}$ Raman results, with the AFM image of \autoref{figure_SEM_AFM} (j), it is clear that MD2 is formed by two crystal grains with similar crystallographic orientations.

In summary, in this section we have demonstrated that a combination of AFM, SEM and Raman spectroscopy is a very powerful approach to address the diamond structural properties, such as crystal stress and grain boundary distribution.

\subsection{\label{sec:SiV_PL_RT}\SiV{} Photoluminescence spectroscopy}

\begin{figure}
	\centering
	\includegraphics[width=1\linewidth]{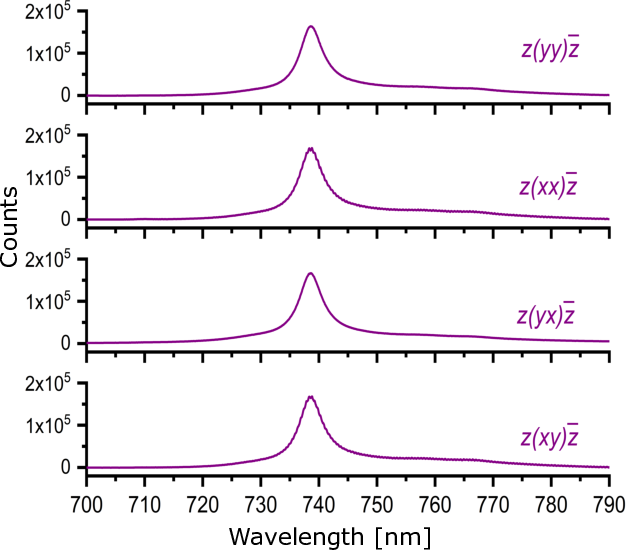}
	\caption{
		Room temperature PL spectra of the \SiV{} centers of MD3 position 4, taken at different polarization configurations: \textit{z(yy)\={z}}, \textit{z(xx)\={z}}, \textit{z(yx)\={z}}, and \textit{z(xy)\={z}}. 
		}
	\label{figure_SiV_PL_RT}
\end{figure}

Photoluminescence (PL) spectra were recorded for numerous MDs under \SI{514}{\nano\meter} excitation, which is known to efficiently excite \SiV{} centres despite being far detuned.
The characteristic \SiV{} PL band was observed in all of the MDs, and the zero-phonon line at \SI{737}{\nano\meter} was investigated in high resolution.
At room temperature there is considerable thermal broadening of this optical transition, and so features were difficult to identify.
However, a range of \SiV{} fluoresence intensities were observed.
\autoref{figure_SiV_PL_RT} reports room temperature, polarization resolved PL measurements taken for MD3 at position 4

\subsection{\label{sec:Montecarlo_SiV}Montecarlo simulation of SiV{} ensambles for size estimation}

\begin{figure}
	\includegraphics[width=1\linewidth]{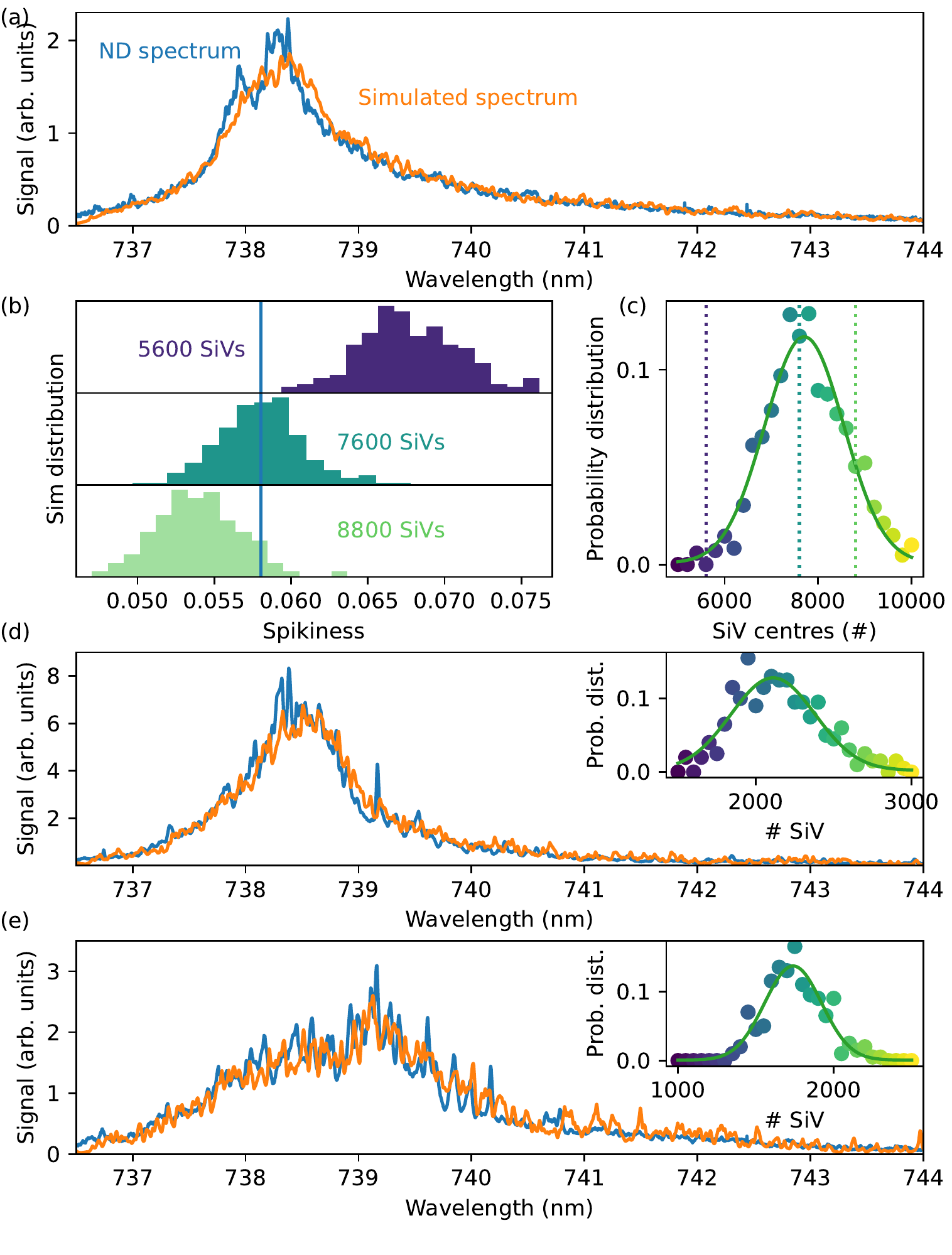}
	\caption{
		Monte Carlo simulated cryogenic spectra to estimate number of \SiV{} centres in MDs.
		(a) The PL spectrum (blue) measured for a MD showed a band-like shape with reproducible spikes superimposed.
		(b) A set of 200 simulated spectra were generated for a given estimate of the total number of \SiV{} centres in the MD.
		The spikiness was calculated for each simulated spectrum, producing the distribution shown in the shaded histogram.
		This was repeated for a range of \SiV{} numbers, and the spikiness distribution from the Monte Carlo simulation compared with the value obtained for the measured spectrum.
		Smaller ensembles (lower numbers of \SiV{} centres) produced spectra with higher spikiness as expected.
		(c) The ``intersection'' of each normalised histogram with the measured spikiness gives a probability distribution over the number of \SiV{} centres.
		An "optimum" simulated spectrum is shown in orange in (a) for easy comparison with the measured data.
		(d) The above process was repeated for other MDs.
		Here the measured spectrum shows more spikiness and the estimated \SiV{} ensemble size is found to be smaller.
		The insert shows the distribution of Monte Carlo simulations over the number of \SiV{} centres.
		(e) A third representative example of this estimation process for yet another MD.
	}
	\label{fig:specsim_num_estimate}
\end{figure}

The sample was cooled to liquid-helium temperatures (below \SI{10}{\kelvin}), and the cryogenic PL spectra exhibited a stable and repeatable pattern of sharp peaks across the ZPL.
Typical examples are shown in \autoref{fig:specsim_num_estimate}.
The \SiV{} ZPL is known to consist of four separate transitions that can be resolved at low temperature, and they can approach the lifetime limited linewidth of about \SI{100}{\mega \hertz} \cite{rogers2014multiple}. 
This is well below the measurement resolution, and it is interpreted that the spiky structure of the measured spectra is due to individual \SiV{} transitions.
For large ensembles of many \SiV{} centres the individual lines average out and the expected spectrum would be a smooth band shaped by the inhomogeneous distribution (predominantly strain variation), whereas for small ensembles of only a few \SiV{} centres the spectrum would be dominated by individual transition lines and a very spiky spectrum would be expected.
This observation provides a way to estimate the number of \SiV{} centres present in the MDs.

The spikiness of each MD spectrum was calculated by first taking a moving-window average of 50 points to obtain the smoothed "band shape". 
The area between the measured spectrum and this smoothed band shape was divided by the area under the band (for normalisation) to produce a quantitative description of the spikiness.
Spikiness values of \SI{5.8e-2}, \SI{9.5e-2}, and  \SI{1.2e-1} were obtained for the three MD spectra shown in \autoref{fig:specsim_num_estimate}.

The \SiV{} transition energies are predominantly influenced by the presence of strain.
Axial strain (along the $\langle111\rangle$ \SiV symmetry axis) shifts the overall transition energy, and transverse strain increases the splitting between the four ZPL components.
At cryogenic temperatures the splitting of the excited-state spin-orbit doublet is enough to cause thermalisation of the excited-state population, and this leads to the higher-energy (blue-shifted) peaks decreasing in intensity relative to the lower-energy peaks.
Thermal broadening of the ZPL components also alters the shape of the \SiV{} spectrum.
Using detailed parameters for these effects from the literature, a \SiV{} spectrum simulator was created capable of producing the expected spectrum given information about temperature and strain.

The running-average smoothed spectral data was used as a probability distribution for randomly sampling the ZPL central position for each SiV centre, and transverse strain was sampled from an exponential distribution to represent the majority of centres having low strain but some high-strain outliers. 
Temperature was set to match the 8K measurement temperature, and each simulated \SiV{} centre was scaled by a brightness factor sampled from a flat distribution (not all colour centres are in the confocal microscope focal spot).
Summing the individual simulated \SiV{} spectra provided Monte Carlo ensemble simulations of the measured MD spectra.
For a chosen number of \SiV centres, 200 simulated ensemble spectra were generated and their spikiness calculated.
The histograms in \autoref{fig:specsim_num_estimate}(b) show examples of the resulting spikiness distributions, which are compared to the single value obtained for the experimental data (vertical line).
As anticipated, increasing the ensemble size (more \SiV{} centres) tends to produce MD spectra with reduced spikiness.

A probability distribution of the number of \SiV{} centres contributing to the measured MD spectrum was obtained by comparing the heights of these histograms at various ensemble sizes with the experimental spikiness.
The MD spectrum featured in \autoref{fig:specsim_num_estimate}(a) produced the distribution illustrated in \autoref{fig:specsim_num_estimate}(c), suggesting this MD contained $7700\pm850$ \SiV{} centres (standard deivation as uncertainty).
This process is summarised in \autoref{fig:specsim_num_estimate}(d)-(e) for two other MDs, which had more spikiness in their spectra and were estimated to have $2100\pm270$ and $1700\pm180$ \SiV centres respectively.

This Monte Carlo simulation of the ensemble spectrum is a novel method for estimating the number of colour centres in medium-sized ensembles.
There is no accessible technique to verify the estimates obtained, but it seems reasonable that MDs grown on silicon substrates would have quite high concentrations of \SiV{} defects.
A few thousand \SiV{} centres per MD corresponds to a few parts-per-million concentrations, which is in line with other \SiV{} samples produced by CVD with silicon etched into the plasma \cite{catledge2011strong}.
\section*{\label{sec:conclusions}Conclusions}
In summary, we demonstrate the fabrication of MDs with large ensemble of \SiV{} centers by MWCVD on Si(001) substrates. 
The morphology and micro-structure of the diamond crystals are fully investigated and correlated to the stress formed at the grain boundary regions. 
We have demonstrated a novel Montecarlo simulation technique to evaluate the number of \SiV{} centers in each MD, based on temperature dependent PL measurements.
The fabrication of MDs with a large density of \SiV{} centers represents a viable solution for numerous applications. 
Bio-marking can benefit from the bright fluorescence within a biological transparency window.
Spatially confined colour centre ensembles are also of interest for fundamental investigations including optical trapping and a recent proposal for strong light-matter coupling in cavities \cite{schutz_ensemble-induced_2020}.
The micro-diamonds presented here are valuable materials for quantum technologies.

\section*{\label{sec:experimental_methods}Experimental Methods}

%
MDs were grown on Si(001) substrates consisting of 15 x 15 mm$^{2}$ chips cleaved from a 4-inch, \SI{500}{\micro\metre} thick, and n-type (1-10 $\Omega$cm) wafer.
Si(001) chips were rinsed in acetone and isopropanol, followed by piranha etching for 15 minutes, and finally cleaned by RF oxygen plasma. 
CVD growth was performed using the MWCVD Seki 6500 reactor at a chamber pressure of \SI{70}{Torr}, with a CH$_{4}$ to H$_{2}$ ratio of 1.5$\%$, at a substrate temperature of 900 $^{\circ}$C for 120 minutes.

%
The SEM analysis was performed using a ZEISS Auriga electron microscope (acceleration voltage \SI{5}{\kilo\volt}, aperture size \SI{30}{\micro\metre}) equipped with a FIB gun. 
The markers were fabricated by FIB using 30 kV Ga$^{+}$ ions and \SI{1}{\nano\ampere} current, they are about \SI{2}{\micro\metre} deep and \SI{30}{\micro\metre} wide, to be easily addressed by optical microscopy.

%
AFM measurements were carried out by a Bruker Dimension ICON SPM microscope at 256 samples/line and \SI{0.3}{\hertz}.
The data are analysed with the open-source software Gwyddion \cite{Necas2011}.

%
Room temperature confocal Raman spectroscopy and \SiV{} PL were performed with a Renishaw inVia system equipped by a green laser (\SI{514}{\nano\metre}), 2400 lines/mm monochromator, 50x 0.9 NA objective, and CCD camera detector.
Polarised measurements are obtained by using a motorised (3 $^{\circ}$/step) rotating $\lambda/2$ waveplate on the laser beam, and a polarizer on the scattered beam. 
%

%
Low temperature PL measurements were performed on a galvo mirror home-built confocal set up, operated using the Qudi software suite \cite{binder2017qudi}.
The silicon substrate was clamped on to the side of an $xzy$ Attocube nano-positioner stack mounted inside a Montana Instruments s200 closed cycle cryostat, with the cold finger at $\sim$ \SI{8}{\kelvin}. 
Off resonant excitation was achieved with a \SI{532}{\nano\meter} Coherent OBIS laser set to \SI{135}{\milli\watt} for spectra measurements. 
PL spectra were taken using a prototype Redback Systems echelle spectrometer, similar in design to the RHEA \cite{feger2014rhea} spectrometer built for astronomical applications.

%
X-ray photoelectron spectroscopy (XPS) analysis was performed using an AXIS Ultra DLD spectrometer (Kratos Analytical Inc., Manchester, UK) with a monochromated Al K$\alpha$ source at a power of 180 $W$ (15 $kV$, 12 $mA$) at a pressure of $\sim$ \SI{1e-9}{\milli\bar}, a hemispherical analyser operating in the fixed analyser transmission mode and the standard aperture (analysis area: \SI{0.3}{\milli\metre} $\times$ \SI{0.7}{\milli\metre}). 
%
%
To obtain more detailed information about chemical structure, high resolution spectra were recorded from individual peaks at \SI{20}{\eV} pass energy. 
The specimen was analysed at an emission angle of 0 $^{\circ}$ as measured from the surface normal. 
Assuming typical values for the electron attenuation length of relevant photoelectrons, the XPS analysis depth (from which 95 \% of the detected signal originates) ranges between 5 to \SI{10}{\nano\metre} for a flat surface.

%
XPS data processing was performed using CasaXPS software version 2.3.15 (Casa Software Ltd., Teignmouth, UK). 
All elements present were identified from survey spectra. 
The atomic concentrations of the detected elements were calculated using integral peak intensities and the sensitivity factors supplied by the manufacturer. 
Binding energies were referenced to the Si 2p$_{3/2}$ peak at \SI{99.4}{\eV} for Si(100).
The accuracy associated with quantitative XPS is about 10 - 15\%.
%

\section*{Acknowledgements}
This work was performed in part at the Melbourne Centre for Nanofabrication (MCN) in the Victorian Node of the Australian National Fabrication Facility (ANFF). 
FI acknowledges the support of the Australian Nanotechnology Network (ANN). 
The authors thank Dr Y. Yao for technical assistance and use of facilities at the Electron Microscope Unit (EMU) within the Mark Wainwright Analytical Centre (MWAC) at UNSW Sydney.
The authors are grateful to Dr T. Gengenbach from CSIRO for XPS measurements and data analysis.
LR is the recipient of an Australian Research Council Discovery Early Career Award (project number DE170101371) funded by the Australian Government.
FI thanks Dr. C. Comte for technical support, Dr. A. Stacey, Dr V. R. Adineh, Dr. L. Brown and Dr. T. van der Laan for fruitful discussions.

\section*{References}

%
\bibliography{library,second_lib}
\bibliographystyle{unsrt}

\section*{Author contributions}
Experiments were conceived by FI, LR, AB, and TV.
FI performed the diamond CVD growth, SEM, FIB and Raman measurements.
MJ, LR and FI performed the spectroscopic measurements.
NHH performed the AFM analysis.
BW and LR developed the Monte Carlo simulation of \SiV{} centers.
The manuscript was written by FI and LR, and all authors discussed the results and commented on the manuscript. 

\section*{Competing financial interests}
The authors declare no competing financial interests.

\end{document}